\documentclass[eqsecnum,twocolumn,showpacs,preprintnumbers,amssymb,aps,pra]{revtex4}
\usepackage{graphicx}
\usepackage{slashbox}
\usepackage{dcolumn}
\usepackage{bm}
\usepackage{latexsym,epsfig}

\usepackage{array}
\usepackage{amstext}

\begin{document}

\title{Parity nonconservation in ytterbium ion}
\vspace*{0.5cm}

\author{B. K. Sahoo \footnote{bijaya@prl.res.in}}
\affiliation{Theoretical Physics Division, Physical Research Laboratory, Ahmedabad-380009, India}

\author{B. P. Das}
\affiliation{Theoretical Astrophysics Group, Indian Institute of Astrophysics, Bangalore-560034, India}

\date{Received date; Accepted date}
\vskip1.0cm

\begin{abstract}
We consider parity nonconservation (PNC) in singly ionized ytterbium (Yb$^+$)
arising from the neutral current weak interaction. We calculate the PNC 
electric dipole transition amplitude ($E1_{PNC}$) and the properties associated
with it using the relativistic coupled-cluster theory.
$E1_{PNC}$ for the
$[4f^{14}] \ ^2 6s \rightarrow [4f^{14}] \ ^2 5d_{3/2}$ transition in Yb$^+$ 
has been evaluated to within an accuracy of 5\%. The improvement of this 
result is possible. It therefore appears that this ion is a promising
candidate for testing the standard model of particle physics.
\end{abstract}

\maketitle

Parity nonconservation (PNC) has been observed in a number of neutral atoms; the latest 
being Yb \cite{ginges,budker}. The combined experimental and theoretical results for Cs PNC
is the most accurate to date \cite{wood}. It is in agreement with the standard model (SM) of
particle physics. The experiment on Cs PNC also reports the observation of the
nuclear anapole moment (NAM) \cite{wood,haxton}.
For more stringent tests of the SM, either the accuracy of the PNC data from Cs 
must be improved or very high precision measurements should be carried out on other candidates. Exploiting
the remarkable advances in laser cooling and single ion trapping techniques, experiments have
been proposed to measure PNC in these systems \cite{fortson,mandal}. Both Ba$^+$ \cite{fortson,mandal} 
and Ra$^+$ \cite{wansbeek,oscar,bijaya1} are already under consideration
for such experiments. Moreover, the PNC nucleon-nucleon coupling constant 
related to the NAM obtained from Cs PNC and nuclear data do not agree
\cite{haxton}, and this clearly calls for further investigation of these 
results. In these circumstances, it would certainly be desirable to explore
PNC in heavy atomic ions \cite{bijaya1}. Apart from  Ba$^+$ and Ra$^+$, we had
also suggested Pb$^+$ as a possible candidate for PNC measurement \cite{bijaya2}.
The study of PNC in the $[4f^{14}] \ ^2 6s \rightarrow [4f^{14}] \ ^2 5d_{3/2}$
transition in Yb$^+$ was suggested by one of us over 
a decade ago \cite{bpdas} and it is currently being experimentally investigated
at the Los Alamos National Laboratory \cite{torgerson}.

Assuming that the PNC light-shift in the above transition in Yb$^+$ can be measured
to sub one per cent accuracy, it would certainly be desirable to determine
the accuracy to which $E1_{PNC}$ can be calculated for this transition.
This knowledge is necessary for choosing a system for investigating atomic PNC.
In alkali atoms and alkali earth-metal ions, the low-lying energy levels
which are the singly excited states from the ground state usually
make the dominant contributions to the $E1_{PNC}$ amplitude. Therefore, a 
typical sum-over-state approach considering only few important singly excited
intermediate states can give a reasonably accurate result. However, Yb$^+$
has a number of doubly excited low-lying intermediate states, and therefore 
it would not be prudent to use a sum-over-states approach for this ion.

In this Rapid Communication, we employ the relativistic coupled-cluster 
(RCC) method to calculate $E1_{PNC}$ using an approach that circumvents
summing over intermediate states \cite{bijaya3}. We also present the
results of our calculations of the excitation energies and the lifetimes for 
the $[4f^{14}] \ ^2 6p_{1/2,3/2}$ and the electric quadrupole moments of the
$[4f^{14}] \ ^2 6s \rightarrow [4f^{14}] \ ^2 5d_{3/2}$ transition, which is 
necessary for evaluating the PNC light-shift associated with this transition 
in Yb$^+$. We analyze
the role of the electron correlation effects in $E1_{PNC}$ and compare them
with those in Ba$^+$ \cite{bijaya3} and Ra$^+$ \cite{wansbeek}.

\begin{figure}[t]
\includegraphics[width=6.8cm]{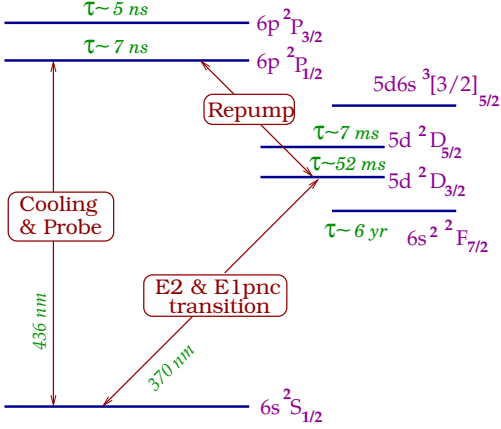}
\caption{(color online) A schematic diagram of energy levels in Yb$^+$ with
transitions shown in red lines that can be induced by different lasers for the 
measurement of parity nonconservation effect. Lifetimes of different states 
are denoted by $\tau$.}
\label{fig1}
\end{figure}

In Fig. \ref{fig1}, we present a diagram of the low-lying energy levels of Yb$^+$
mentioning lifetimes of the important states and transitions that can be induced
by lasers for the measurement of PNC. As shown in the
figure, the lifetimes of the $5d_{3/2,5/2}$ states are fairly large
\cite{gerz,yu,taylor}.

The largest contribution to $E1_{PNC}$ in the $[4f^{14}] \ ^2 6s \rightarrow 
[4f^{14}] \ ^2 5d_{3/2}$ transition comes from the nuclear spin  independent
(NSI) neutral weak current interaction between the electrons and the nucleus
due to the exchange of $Z_0$ boson. The Hamiltonian for this 
interaction is given by
\begin{eqnarray}
H_{PNC} = \frac{G_F}{2\sqrt{2}} Q_W \gamma_5 \rho_{nuc}(r),
\end{eqnarray}
where $G_F$($=2.219 \times 10^{-14}$ in au) is the Fermi constant that 
quantifies the strength of the weak interaction, $\rho_{nuc}$ is the nuclear
density which is evaluated using the Fermi distribution, $\gamma_5$ is the 
Dirac matrix, $N$ is the neutron number of the atomic system and $Q_W$
is known as the weak nuclear charge. As the strength of the weak
interaction is very small compared to the electromagnetic interactions
in an atomic system, it can be treated as a first order perturbation.
The wave function of an atomic state with a single valence electron $v$ can 
be expressed as
\begin{eqnarray}
\vert \Psi_v \rangle = \vert \Psi_v^{(0)} \rangle + G_F \vert \Psi_v^{(1)} \rangle,
\end{eqnarray}
with $G_F$ representing the perturbed parameter, $\vert \Psi_v^{(0)}\rangle$ 
and $\vert \Psi_v^{(1)} \rangle$ are the wave functions corresponding to the
the  Dirac-Coulomb Hamiltonian ($H_{DC}$) and its first order correction due to
the PNC weak interaction, respectively. In our RCC approach, we write
\begin{eqnarray}
\vert \Psi_v \rangle &=& e^T [ 1+S_v ] \vert \Phi_v \rangle,
\end{eqnarray}
where $\vert \Phi_v\rangle$ is the Dirac-Fock (DF) wave function obtained by 
appending the valence electron $v$ to the closed-shell configuration reference
state $\vert \Phi_0\rangle$ that is $[4f^{14}]$ in the present system and $T$
and $S_v$ are the   
core-virtual and valence-virtual excitation operators, respectively. In this
work, we have considered all possible single and double excitations (CCSD 
method) by taking the operators
\begin{eqnarray}
T = T_1 + T_2
 \ \ \ \text{and} \ \ \ S_{v} = S_{1v} + S_{2v},
\end{eqnarray}
where subscripts $1$ and $2$ stand for the single and double excitations, 
respectively. We also consider contributions from triple excitations involving the
valence electron by defining the following operator
\begin{eqnarray}
 S_{3v} &=& \frac {H_{DC} T_2 + H_{DC} S_{2v}}{\varepsilon_p + \varepsilon_q + \varepsilon_r - \varepsilon_v - \varepsilon_a - \varepsilon_b},
\end{eqnarray}
where $\varepsilon$ represents the orbital energy and $p,q,r$ and $a,b$ represent 
the virtual and occupied orbitals, respectively. When $S_{3v}$ is used
only to improve the CCSD wave functions, it is referred to as the CCSD(T) method and when 
its contributions are also estimated then it is known as lo-CCSDvT method 
\cite{bijaya4}. Further, we separate the $T$ and $S_v$ operators to represent 
the unperturbed and perturbed wave functions as
\begin{eqnarray}
T = T^{(0)} + G_F T^{(1)} \ \ \ \text{and} \ \ \ S_{v} = S_{v}^{(0)} + G_F S_{v}^{(1)},
\end{eqnarray}
where the superscripts $0$ and $1$ denote unperturbed and perturbed operators,
respectively. In an explicit form, we can write
\begin{eqnarray}
\vert \Psi_v^{(0)} \rangle &=& e^{T^{(0)}} [ 1 + S_v^{(0)} ] \vert\Phi_v\rangle \\
\text{and} \ \ 
\vert \Psi_v^{(1)} \rangle &=&  e^{T^{(0)}} [ T^{(1)} (1 + S_v^{(0)} ) + S_v^{(1)} ] \vert\Phi_v\rangle. \ \ \ \ \
\end{eqnarray}
The RCC equations for the correlation energy and the cluster amplitudes 
equations for both the unperturbed and perturbed wave functions have been given
in some of our earlier papers (for example see \cite{bijaya3,bijaya4}).

An atomic property can be determined in the RCC framework by evaluating the 
matrix element of an operator $O$ corresponding to that property
\begin{eqnarray}
\langle O \rangle_{fi} &=& \frac{ \langle \Phi_f \vert [ 1+ S_f^{(0)^\dagger} ] {\cal O}_0 [ 1+ S_i^{(0)} ] \vert \Phi_i \rangle } { \sqrt{ {\cal N}_f {\cal N}_i } } ,
\end{eqnarray}
where ${\cal O}_0=e^{{T^{(0)}}^\dagger} O e^{T^{(0)}}$ and ${\cal N}_v = \langle \Phi_v \vert { [ 1 + S_v^{(0)^{\dagger}} ] \cal N}_0 [ 1+ S_v^{(0)} ] \vert \Phi_v \rangle$ with ${\cal N}_0=e^{{T^{(0)}}^\dagger} e^{T^{(0)}}$. Expectation values are determined by taking $f=i$.

The expression for $E1_{PNC}$ between $\vert \Psi_f \rangle$ and $\vert \Psi_i
\rangle$ after keeping terms up to linear in $G_F$ is given by
\begin{eqnarray}
E1_{PNC} &=& G_F \frac {\langle \Psi_f^{(0)} \vert D \vert \Psi_i^{(1)} \rangle + \langle \Psi_f^{(1)} \vert D \vert \Psi_i^{(0)} \rangle } {\sqrt {\langle \Psi_f^{(0)} \vert \Psi_f^{(0)} \rangle \langle \Psi_i^{(0)} \vert \Psi_i^{(0)} \rangle } }, \ \ \
\end{eqnarray}
for the electric dipole (E1) operator $D$, which in the RCC approach leads to
\begin{eqnarray}
\frac{E1_{PNC}}{G_F} = \frac{ \langle \Phi_f \vert [1+ S_f^{(0)^\dagger}] {\cal D}_0 [ T^{(1)} (1+ S_i^{(0)}) + S_i^{(1)} ] \vert \Phi_i \rangle } { \sqrt{ {\cal N}_f {\cal N}_i } }  && \ \ \ \nonumber \\
 + \frac{ \langle \Phi_f \vert [ S_f^{(1)^\dagger} + (1+ S_f^{(0)^\dagger}) T^{(1)^\dagger} ] {\cal D}_0 [ 1+ S_i^{(0)}] \vert \Phi_i \rangle } { \sqrt{ {\cal N}_f {\cal N}_i } }. && \nonumber
\end{eqnarray}

\begin{table}[t]
\caption{Comparison of different properties in $^{171}$Yb$^+$ from the present and other works (given up to second decimal place). Uncertainties are given inside the parentheses.}
\begin{ruledtabular}
\begin{center}
\begin{tabular}{cccc}
Properties           & This work & Others & Experiment \\
\hline
                  &     &  & \\
Transitions & \multicolumn{3}{c}{Excitation energies (in cm$^{-1}$)} \\ \cline{2-4}
$6s_{1/2} \rightarrow 6p_{1/2}$ & 28109(1000) & 28048$^a$ & 27061.82$^b$\\
$6s_{1/2} \rightarrow 6p_{3/2}$ & 31604(800) & 31411$^a$ & 30392.23$^b$ \\
$6s_{1/2} \rightarrow 7p_{1/2}$ & 63518(200) & 63227$^a$ & 63706.25$^b$ \\
$6p_{1/2} \rightarrow 5d_{3/2}$ & 4879(800) & 6810$^a$  & 4101.02$^b$ \\
$6p_{3/2} \rightarrow 5d_{3/2}$ & 8375(950) & 10173$^a$ & 7431.43$^b$ \\
$6p_{3/2} \rightarrow 5d_{5/2}$ & 7015(1000) & 8962$^a$  & 6059.54$^b$ \\
$7p_{3/2} \rightarrow 5d_{3/2}$ & 41502(1500) & 43175$^a$ & 42633.30$^b$ \\
                 &     &  & \\
 & \multicolumn{3}{c}{E1 matrix elements (in au)} \\ \cline{2-4}
$6s_{1/2} \rightarrow 6p_{1/2}$ & 2.72(1) & 2.68/2.73$^a$, 2.76$^c$ &  \\
$6s_{1/2} \rightarrow 6p_{3/2}$ & 3.83(1) & 3.77/3.84$^a$, 3.87$^c$  &  \\
$6s_{1/2} \rightarrow 7p_{3/2}$ & 0.18(2) & 0.15/0.04$^a$  & \\
$6p_{1/2} \rightarrow 5d_{3/2}$ & 3.06(2) & 2.97/3.78$^a$, 3.19$^c$ & \\
$6p_{3/2} \rightarrow 5d_{3/2}$ & 1.35(2) & 1.31/1.55$^a$, 1.40$^c$ & \\
$6p_{3/2} \rightarrow 5d_{5/2}$ & 4.23(3) & 4.12/4.77$^a$, 4.39$^c$ & \\
$7p_{1/2} \rightarrow 5d_{3/2}$ & 0.17(2) & 0.08/0.12$^a$  & \\
                  &     &  & \\
 & \multicolumn{3}{c}{E2 matrix elements (in au)} \\ \cline{2-4}
$6s_{1/2} \rightarrow 5d_{3/2}$ & 10.2(5) &  & \\
$6s_{1/2} \rightarrow 5d_{5/2}$ & 12.9(7) & & 12(1) \\
                  &     &  & \\
States             & \multicolumn{3}{c}{$A_{hyf}$ constants (in MHz)} \\ \cline{2-4}
$6s_{1/2}$   & 13332(1000) & 13172$^a$, 12730$^d$ & 12645(2)$^d$ \\
$6p_{1/2}$   & 2516(400) & 2350$^a$, 2317$^d$ & 2104.9(1.3)$^d$ \\
$6p_{3/2}$   &  322(20) & 311.5$^a$, 391$^d$ &  877(20)$^e$ \\
$7p_{1/2}$   &  861(50) & 807.4$^a$ &  \\
$7p_{3/2}$  & 123(15) & 110.8$^a$ &  \\
$5d_{3/2}$   & 447(20)  & 400.48$^f$ &  430(43)$^g$ \\
$5d_{5/2}$   & $-48(15)$  & $-12.58^f$ &  $-63.6(7)^h$ \\
\end{tabular}
\end{center}
\end{ruledtabular}
\label{tab1}
References: 
$^a$\cite{safronova}; $^b$\cite{nist}; $^c$\cite{mani1}; $^d$\cite{martensson}; $^e$\cite{berends}; $^f$\cite{itano}; $^g$\cite{engelke}; $^h$\cite{roberts}.
\end{table}

In Table \ref{tab1}, we present results for different properties using our
CCSD(T) method. The results for the appropriate properties can 
be used for estimating the uncertainty of our $E1_{PNC}$ 
result. We also compare our results with previous calculations and experimental 
results wherever possible. We analyze the results briefly below.

{\it Basis functions:} We have considered up to 17 $s$, $p$,
$d$ and $f$ and 14 $g$ orbitals using
Gaussian type orbital basis functions. A Fermi-nuclear charge distribution
is used to determine the nuclear potential and density.

{\it Excitation energies:} We have given excitation energies (EEs) in Table 
\ref{tab1} corresponding to the transitions that are important in obtaining an 
accurate value for $E1_{PNC}$ result and the lifetimes of excited states. 
Among them EE of the $6s_{1/2} \rightarrow 6p_{1/2}$ transition contribute 
significantly to the $E1_{PNC}$ result which is obtained to an accuracy of 
around 4\% 
in this work. All our results are compared with MBPT(3) results of Safronova
and Safronova \cite{safronova} and also with experimental values \cite{nist}.

{\it E1 matrix elements:} It is important to find the accuracy of the 
E1 matrix elements that contribute to $E1_{PNC}$. We have presented the
important E1 matrix elements in the above table. Our results compare reasonably
well with previous calculations \cite{safronova,mani}. In \cite{safronova}, two
different approaches were considered to obtain these results which differ from
each other substantially. We have compared our results with Mani and Angom
(their published results \cite{mani} have been corrected \cite{mani1}).
We combine our E1 matrix elements with the experimental wavelengths
 to determine the lifetimes of the important intermediate states; $6p \ 
^2P_{1/2}$ and $6p \ ^2P_{3/2}$ states and compare them with the measured 
values in order to test their accuracies. Using our calculated E2 and M1 amplitudes; i.e.  $20.70$ au and $1.15$ au,
respectively for the $6p_{3/2} \rightarrow 6p_{1/2}$ transition, we find lifetimes of the $6p \ ^2P_{1/2}$ 
and $6p \ ^2P_{3/2}$ states as 6.70 ns and 4.75 ns, which agree reasonably with
the experimental results $7.1(4)$ ns and $5.5(3)$ ns of Blagoev et al 
\cite{blagoev}, respectively. Also, the branching ratio of the decay rate of the 
$6p \ ^2P_{1/2}$ state to the $5d \ ^2D_{3/2}$ state has been reported as 0.0483 with 9\%
uncertainty \cite{yu} and it agrees well with our result 0.0439.

{\it E2 matrix elements:} We have also determined the E2 matrix elements of the
$6s_{1/2} \rightarrow 5d_{3/2}$ and $6s_{1/2} \rightarrow 5d_{5/2}$ transitions
which are needed in evaluating the PNC light shift. No other results are available for
comparison. Taylor et al have measured the lifetimes of the $5d \ ^2D_{5/2}$
state as 7.2(3) ms with a branching ratio to the $^2F_{7/2}$ state of 0.83(3)
\cite{taylor}. From their measurements, we estimate the E2 matrix element of 
the $6s_{1/2} \rightarrow 5d_{5/2}$ transition as 12(1) au and our result
is in agreement with that. Therefore, we expect that our E2 transition amplitude
for the $6s_{1/2} \rightarrow 5d_{3/2}$ transition will also be of 
similar accuracy.

{\it Hyperfine structure constants:} We have presented the magnetic dipole 
hyperfine structure constants ($A_{hyf}$) of the low-lying singly excited states
of $^{171}$Yb$^+$. They are also compared with other calculations and experimental 
results. Our results are comparatively larger than the experimental 
values, but all the measurements are relatively old. A large disagreement 
between experimental and all theoretical results of the $6p \ ^2P_{3/2}$ state 
calls for verification of these results using new experimental techniques. However,
the accuracies of our results for obtaining a reasonable estimate of
$E1_{PNC}$ are currently acceptable. For high precision results, it would be necessary to
improve the accuracies of the calculations.

\begin{table}[t]
\caption{$E1_{PNC}$ results in Yb$^+$ in $10^{-11} iea_0 (-Q_W/N)$ using different approximations.}
\begin{center}
\begin{tabular}{c|cccc}
\hline
\hline
\backslashbox{Result}{Method} & DF & CCSD & CCSD(T) & lo-CCSDvT \\
\hline
 & & \\
$E1_{PNC}$ & $6.5311$ & $8.5126$ & $8.4702$ & $8.4704$ \\
\hline
\end{tabular}
\end{center}
\label{tab2}
\end{table}

{\it $E1_{PNC}$ result:} We turn here to our $E1_{PNC}$ calculation of the
$6s_{1/2} \rightarrow 5d_{3/2}$ transition where the correlation 
effects seem to be significant. It can be seen from Table 
\ref{tab2} that the $E1_{PNC}$ result increases by 30\% and 29\% at the CCSD
and CCSD(T) levels, respectively. So there are some cancellations between
the correlation contributions coming via the CCSD and CCSD(T) methods.

\begin{table}[t]
\caption{Contributions (Contr.) to $E1_{PNC}$ result (in $\times 10^{-11} iea_0 (-Q_W/N)$) in Yb$^+$ at the DF and CCSD(T) methods as initial ($i=6s \ ^2S_{1/2}$) and final ($f=5d \ ^2D_{3/2}$) perturbed states. Results given as $Norm$ and $Others$ refer to contributions from normalization of the wave functions and RCC terms those are not mentioned explicitly, respectively. Subscripts $c$ and $vir$ represent contributions from core and virtual orbitals, respectively.}
\begin{tabular}{llclc}
\hline
\hline
Method & Initial state & Contr. & Final state & Contr. \\
 & & \\
DF & $(H_{PNC}D)_{c}$ & $0.6441$ & $(DH_{PNC})_{c}$ & $-3.3\times10^{-7}$ \\
 & $(D H_{PNC})_{vir}$ & $5.8870$ & $(H_{PNC} D)_{vir}$ & $-4.3\times10^{-5}$ \\
\hline
 & & \\
RCC & $ T_1^{(1)\dagger} {\cal D}_0 $ & $0.6745$ &  ${\cal D}_0 T_1^{(1)} $ & $0.0070$ \\
 & ${\cal D}_0 S_{1i}^{(1)}$ & $7.1973$ & $S_{1f}^{(1)\dagger} {\cal D}_0 $ & $1.3424$\\
 & ${\cal D}_0 S_{2i}^{(1)}$ & $-0.2929$ & $S_{2f}^{(1)\dagger} {\cal D}_0 $ & $-0.2022$  \\
 & $S_{1f}^{(0)\dagger} {\cal D}_0 S_{1i}^{(1)}$ & $0.5115$ & $S_{1f}^{(1)\dagger} {\cal D}_0 S_{1i}^{(0)} $ & $-0.0788$ \\
 & $S_{2f}^{(0)\dagger} {\cal D}_0 S_{1i}^{(1)}$ & $-0.3682$ & $S_{1f}^{(1)\dagger} {\cal D}_0  S_{2i}^{(0)} $ & $-0.0617$ \\
 & $S_{1f}^{(0)\dagger} {\cal D}_0 S_{2i}^{(1)}$ & $0.0004$ & $S_{2f}^{(1)\dagger} {\cal D}_0 S_{1i}^{(0)} $ & $-0.0165$ \\
 & $S_{2f}^{(0)\dagger} {\cal D}_0 S_{2i}^{(1)}$ & $0.0271$ & $S_{2f}^{(1)\dagger} {\cal D}_0  S_{2i}^{(0)} $ & $0.0096$ \\
 & $Others$ & $-0.6261$ &  & 0.7078 \\
 & $Norm$ & -0.2911 & & $-0.0670$ \\
\hline
\hline
\end{tabular}
\label{tab3}
\end{table}

We now proceed to discuss the trends of the correlation effects by considering 
the contributions from different RCC terms. In Table \ref{tab3}, we give 
contributions from the DF and individual CCSD(T) terms. By relating the
important CCSD(T) terms to their corresponding lower order MBPT
counterparts, we can identify the correlations effects that are
significant. As seen in Table \ref{tab3}, three different classes of RCC terms
are of crucial importance in arriving at the final result. The
 ${\cal D}_0 T_1^{(1)}$ term represents a sub class of core-valence 
correlations, whose contribution in the present case is slightly larger than
DF result that is denoted as $(D H_{PNC})_{c}$.
As we have previously discussed \cite{bijaya3,bijaya4}, ${\cal D}_0 S_{1v}^{(1)}$ effectively considers contributions from singly excited
states with $(D H_{PNC})_{vir}$ as its leading term and involves
contributions from important pair-correlation and a class of core-polarization
effects. Clearly this contribution through the perturbed $5d \
^2D_{3/2}$ state at the DF level is negligible, but after including electron
correlation effects, its contribution becomes large. The other important 
contributions coming via ${\cal D}_0 S_{2v}^{(1)}$ are from the core polarization effects
involving doubly excited states such as $4f^{13}
\ (^2F_{5/2}) \ 6s^2$ and $4f^{13} \ (^2F_{7/2}) \ 5d6s$. A sizable amount of contribution comes
collectively from certain higher order RCC terms given as $Others$.
Contributions from the normalization of the wave
functions (given as $Norm$) are also not small.

A detailed comparative study of the trends of
the correlation effects in $E1_{PNC}$ amplitudes of the
$s_{1/2} \rightarrow 5d_{3/2}$ transitions in Yb$^+$ and similar
transitions in Ba$^+$ and Ra$^+$ will be reported elsewhere. However,
there is an important point that we would like to highlight here.
There is a cancellation between
${\cal D}_0 S_{1v}^{(1)}$ and its complex conjugate ($cc$) term in Ba$^+$ and 
Ra$^+$, but in contrast in the present case it adds up. Therefore, the net
correlation contribution is larger in Yb$^+$ than other two ions.

Analysing results from Table \ref{tab3}, we find that the leading contributions from ${\cal D}_0 T^{(1)}+cc$ through the core electrons,
$[{\cal D}_0 S_{1i}^{(1)} + S_{1f}^{(0)\dagger}{\cal D}_0 S_{1i}^{(1)}] + cc$ as in the form of the single excitations of the valence electrons and the double excitation terms $[{\cal D}_0 S_{2i}^{(1)} + S_{1f}^{(0)\dagger}{\cal D}_0 S_{2i}^{(1)}+ S_{2f}^{(0)\dagger}{\cal D}_0 S_{2i}^{(1)}] + cc$
are around 8\%, 101\% and 9\%
(with opposite sign), respectively. The uncertainties in the contributions
from the core and doubly excited states are taken as the differences
of the CCSD and lo-CCSD(T) results. Uncertainties from the singly excited
states are estimated from the errors of various physical quantities
which are given in Table \ref{tab1} and taking into account their
contributions with appropriate weights. With those uncertainties, we estimate
the $E1_{PNC}$ amplitude for the $6s \ ^2S_{1/2} \rightarrow 5d \ ^2D_{3/2}$
transition in Yb$^+$ to be $8.5(5)\times 10^{-11} \ iG_F ea_0 (-Q_W/N)$.
The results can be improved by including the triple and quadruple
hole-particle excitations in the framework of general-order RCC theory \cite{kallay}.

In conclusion,Yb$^+$ appears to be a promising candidate for the study 
of atomic PNC. By employing the RCC method in the singles, doubles and
partial triples approximation, we have estimated various atomic
properties including the $E1_{PNC}$ 
amplitude. Our preliminary investigation suggests that it is
feasible to obtain theoretical results in this system with 
high accuracy after suitable modifications to our present method
which has yielded a result that is about 5\% accurate. 
A different trend for correlation effects is found in the present case
than Ba$^+$ and Ra$^+$; the two ions that were 
previously studied using the same method.

We thank J. R. Torgerson for communicating with us. The computation of the 
results presented in this work were obtained using PRL HPC 3TFLOP cluster.

\end{document}